\documentclass[pre,showpacs,floatfix,twocolumn,superscriptaddress]{revtex4}
\usepackage{amsmath}
\usepackage{amssymb}
\usepackage{latexsym}
\usepackage{multirow}
\usepackage{color}
\usepackage[hidelinks]{hyperref}
\usepackage{mathtools}
\usepackage{comment}
\usepackage{graphicx}
\usepackage{bm}

\begin{document}

\title{Flocking of two unfriendly species: The two-species Vicsek model}

\author{Swarnajit Chatterjee}
\email{swarnajit.chatterjee@uni-saarland.de}
\affiliation{Center for Biophysics and Department of Theoretical Physics, Saarland University, 66123 Saarbr{\"u}cken, Germany.}

\author{Matthieu Mangeat}
\email{mangeat@lusi.uni-sb.de}
\affiliation{Center for Biophysics and Department of Theoretical Physics, Saarland University, 66123 Saarbr{\"u}cken, Germany.}

\author{Chul-Ung Woo}
\affiliation{Department of Physics, University of Seoul, Seoul 02504, Korea.}

\author{Heiko Rieger}
\email{heiko.rieger@uni-saarland.de}
\affiliation{Center for Biophysics and Department of Theoretical Physics, Saarland University, 66123 Saarbr{\"u}cken, Germany.}
\affiliation{Leibniz-Institute for New Materials INM, 66123 Saarbr{\"u}cken, Germany}

\author{Jae Dong Noh}
\email{jdnoh@uos.ac.kr}
\affiliation{Department of Physics, University of Seoul, Seoul 02504, Korea.}

\date{\today}

\begin{abstract}
We consider the two-species Vicsek model (TSVM) consisting of two kinds of self-propelled particles, A and B, that tend to align with particles from the same species and to anti-align with the other. The model shows a flocking transition that is reminiscent of the original Vicsek model: it has a liquid-gas phase transition and displays micro-phase separation in the coexistence region where multiple dense liquid bands propagate in a gaseous background. The interesting features of the TSVM are the existence of two kinds of bands, one composed of mainly A-particles and one mainly of B-particles and the appearance of two dynamical states in the coexistence region: the PF (parallel flocking) state in which all bands of the two species propagate in the same direction, and the APF (anti-parallel flocking) state in which the bands of species A and species B move in opposite directions. When PF and APF states exist in the low-density part of the coexistence region they perform stochastic transitions from one to the other.  The system size dependence of the transition frequency and dwell times shows a pronounced crossover that is determined by the ratio of the band  width and the longitudinal system size. Our work paves the way for studying multi-species flocking models with heterogeneous alignment interactions.
\end{abstract}

\maketitle

\section{Introduction}
\label{s1}

Active matter is a class of natural or synthetic non-equilibrium systems composed of a large number of agents that consume energy in order to move or to exert mechanical forces \cite{AM-Reviews1,AM-Reviews2,AM-Reviews3,AM-Reviews4}. An assembly of active particles behaves in complex ways and shows collective effects such as the emergence of coherent motion of large clusters or flocks. Flocking is observable on a wide range of scales, from mammalian herds, fish schools, and sterling flocks to amoeba and bacteria colonies, to the cooperative behavior of molecular motors in living cells or in vitro environments. Physically flocking of self-propelled particles is equivalent to the appearance of long-range order and thus related to a spontaneous breaking of a symmetry of the system \cite{VM,vicsek97,toner-tu,chate-lro}.

The Vicsek model (VM)~\cite{VM} was introduced as the simplest and prototypical model that shows a flocking transition, where point particles with an $O(2)$ rotational symmetry tend to align with the average direction of motion of their neighbors while moving at a fixed speed and being submitted to some noise. The VM has a phase transition to a kinetic, swarm-like phase when it approaches a critical value of the noise parameter. By varying the noise level in the system, the density of the individuals, and the individual radius, the Vicsek model can be switched from a gas-like phase, in which the individuals move almost independently of each other, to a swarming phase, in which individuals self-organize in clusters. Although the VM displays a transition from 
a disordered low density/high noise to a high density/low noise phase, 
it was shown by Solon and collaborators~\cite{SolonVM} that the VM is best understood in terms of a liquid-gas transition with micro-phase separation in the coexistence region. 

Complex systems are typically heterogeneous as individuals vary in their
properties, their response to the external environment and to each
other~\cite{hetero}. In particular, many biological systems that show flocking involve self-propelled particles with heterogeneous interactions (e.g. bacterial collectives typically consist of multiple species), which motivates the study of  populations with multiple species. 

In Ref. \cite{mixed-species-Ariel1}, the collective dynamics of mixed swarming bacterial populations composed of cells of one species but different phenotype, specifically with different aspect ratios (“length”) was experimentally studied. In contrast to the homogeneous system the mixture did not show macroscopic phase separation, but locally long cells acted as nucleation cites, around which aggregates of short, rapidly moving cells can form, resulting in enhanced swarming speeds.

Similarily in Ref. \cite{population-segregation}, a population of single species bacteria, Escherichia coli, with antibiotics induced heterogeneous motility was studied, which was found to promote the spatial segregation of subpopulations via a dynamic motility selection process. Contrastingly, in Ref. \cite{mixed-species-Ariel2} a mixture of two different swarming bacterial species was studied and it was found that the mixed population swarms together well and that the fraction between the species determines all dynamical scales—from the microscopic (e.g., speed distribution), to the mesoscopic (vortex size), and macroscopic (colony structure and size). 

Theoretically various aspects of heterogeneous systems of self-propelled agents have been investigated.  Examples include particles and agents with varying velocities \cite{mixed-velocities}, noise sensitivity \cite{mixed-noise}, sensitivity to external cues \cite{mixed-external}, and particle-to-particle interactions \cite{mixed-interactions}. Different self-propelled particle species were also analyzed in predator-prey scenarios \cite{pp1,pp2} and in the context of a non-reciprocal interaction~\cite{fruchart}. 

One step further one could for instance ask, what happens when two unfriendly species, each of which tries to avoid the other one, are forced to encounter in a confined environment: (a) Does a collective behavior emerge in this multi-species system? (b) If then, how the two different species move? (c) What is the impact of heterogeneity on the order-disorder phase transition? (d) What is the spatial structure of the ordered phase? etc. 
Here we try to address these questions by focusing on the effect of alignment interactions between different particle species, similar to what has been done in Ref. \cite{Menzel}. The latter work considered a binary mixture of self-propelled particles described by Langevin equations and specific interaction potentials leading to parallel, anti-parallel or perpendicular alignment and a variety of collective motion patterns was found. 
To allow for a detailed quantitative analysis 
of all the emerging phases and phase diagrams, including dynamical phenomena, 
we focus here on a more simplified model, closely related to the original Vicsek model but equipped with two particle species: the two-species Vicsek model (TSVM) with intra-species alignment and inter-species anti-alignment. 

We will show that the TSVM has a flocking transition reminiscent 
of the original VM, but shows different dynamical states (parallel
and anti-parallel flocking, PF and APF) in parts 
of the coexistence region plus size-dependent transitions between them 
- and a liquid phase in which the two species move in opposite directions.
The paper is organized as follows: the model is introduced in Sec.~\ref{s2},
the emerging collective motion and the phase diagrams are presented in
Sec.~\ref{s3}, the dynamics of the stochastic transition between the 
PF and APF states is analyzed in Sec.~\ref{s4}, and Sec.~\ref{s5} 
summarizes and discusses our findings.

\section{Model}
\label{s2}

The two-species flocking model (TSVM) that we consider here is based on the
original VM, consisting of self-propelled point-like particles 
moving in two dimensions with alignment interactions, and comprises two
different kinds of particles, two “species” A and B.
As a first step in studying multi-species flocking, we assume that
each particle tends to align with particles of the same species and
anti-aligns with particles of the other species.

Formally there are $N_A$~($N_B$) active particles of species A~(B) in
a two-dimensional (2D) rectangular geometry of size $L_x \times L_y$ with
periodic boundary conditions. Each one carries an off-lattice position
vector $\bm{r}_i=(x_i,y_i)$, a unit orientation vector
$\bm{\sigma}_i=(\cos \theta_i,\sin \theta_i)$ with an orientation angle
$\theta_i$ representing 
its self-propulsion direction, and a static Ising-like spin variable 
$s_i=\pm 1$ signifying the 
species it belongs to~($s_i=+1$ for an A particle and $s_i=-1$ for a B
particle). Particles are self-propelled and move at a constant speed $v_0$ in
the direction of the orientation vector. 
In this paper we focus on the equal population case 
$N_A=N_B$, if not stated otherwise. The total number of particles 
is denoted as $N$.

At each discrete time step $\Delta t$, a particle $i$ interacts with
neighboring particles within a distance $R_0$, denoted as $\mathcal{N}_i$,
and evolves its orientation and position in the following way:
\begin{gather}
\theta_i^{t+\Delta t}= \bar{\theta}_i^t + \eta \xi_i^t \,
,\label{sigma} \\
\bm{r}_i^{t+\Delta t}=\bm{r}_i^t+ v_0 \bm{\sigma}_i^{t+\Delta t} \Delta t \, ,\label{r} 
\end{gather}
where $\bar{\theta}_i^t$ is the orientation angle of a spin-weighted sum
\begin{equation}
\label{sigma2}
\bar{\bm{\sigma}}_i^t = \sum_{j\in \mathcal{N}_i} s_i s_j \bm{\sigma}_j^t,
\end{equation}
of orientation vectors of neighboring particles, and $\xi_i^t$ is a scalar 
noise distributed uniformly in $[-\pi,\pi]$ satisfying 
\begin{equation}
    \langle \xi_i^t \rangle=0, \quad \langle \xi_i^t \xi_j^{s} \rangle \sim
    \delta_{ts} \delta_{ij}\, .
\end{equation}
The noise strength is controlled by the parameter $\eta$.
Due to the spin-dependent factor $s_i s_j$, a particle tends to flock 
together with particles of the same species~($s_i s_j=1$) and
anti-flock with those of the other species~($s_i s_j = -1$). The model can
be generalized to a multi-species model by introducing a general spin
variable $\bm{s}$ other than the Ising variable and a suitable interaction factor 
$f(\bm{s}_i, \bm{s}_j)$, but we restrict ourselves to the simplest
case here.

\begin{figure}[t]
    \includegraphics[width=\columnwidth]{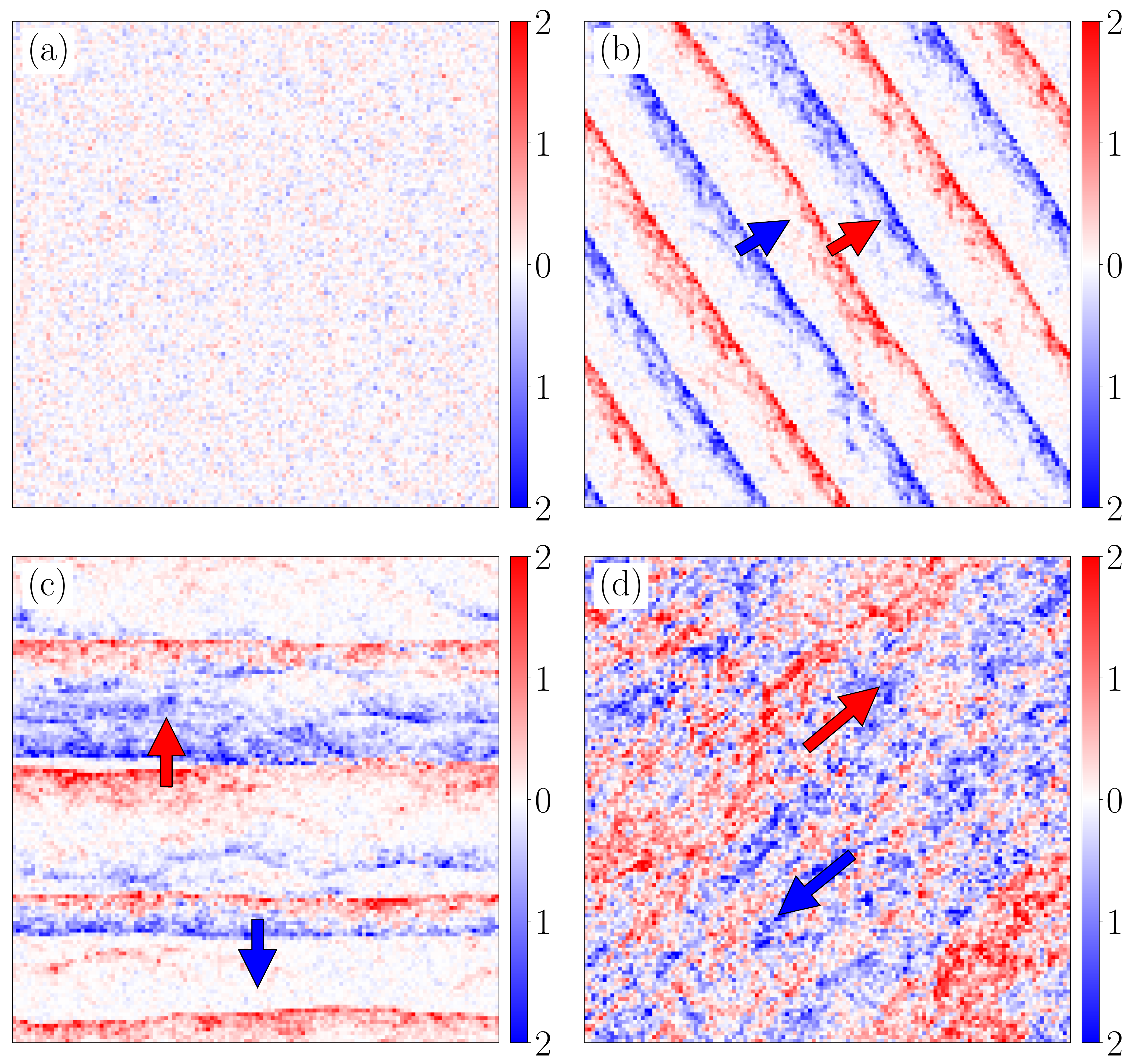}
\caption{Typical configurations of the TSVM (a) in a gas state at $(\rho,\eta)=(0.5,0.3)$, (b) in a parallel flocking (PF) state at $(\rho,\eta)=(0.5,0.24)$, (c) in an anti-parallel (APF) flocking state at $(\rho,\eta)=(0.5,0.2)$, and (d) in a liquid state at $(\rho,\eta)=(2.0,0.2)$ (always APF). Configurations are for a square domain of size $2048\times 2048$ with $v_0=0.5$. Coarse-grained local densities of A- and B-particles are represented with red and blue pixels, respectively, color coded according to the color bar. In (b,c,d) particle species moves collectively in the direction indicated by an arrow.}
\label{fig1}
\end{figure}

Model parameters are the particle number density $\rho = N/L_xL_y$, noise strength $\eta$, and velocity modulus $v_0$. 
The unit of space $R_0$ and time $\Delta t$ is set to be unity, $R_0=\Delta t=1$. The particle number density of each species, which will be denoted by $\rho_0$, is given by $\rho_0 = \rho/2$.

We performed numerical simulations of the stochastic process with parallel update. We consider random initial conditions by assigning random position and orientation to each particle. After the initialization, we let the system evolve under various control parameters for $t_{\rm eq} \sim 10^5$ to reach the steady-state and measure various quantities until the maximum simulation time $t_{\rm max} \sim 10^6$.
Fig.~\ref{fig1} shows typical steady state configurations at various model parameter values. These snapshots suggest that the system exists in distinct phases, which will be characterized in the following section.

\section{Collective motion and phase diagrams}
\label{s3}

We find that the TSVM undergoes a liquid-gas phase transition with an
intermediate phase coexistence region.
In the gas phase~(low density and high noise), particles are distributed uniformly and move incoherently (c.f. Fig.~\ref{fig1}(a)). In the liquid phase (high density and low noise), each particle species performs collective flocking with giant density fluctuations (c.f. Fig.~\ref{fig1}(d)). In the coexistence region each species forms an array of liquid bands traveling coherently in a gaseous background (see Fig.~\ref{fig1}(b) and (c)). These phenomena are reminiscent of the liquid-gas phase transition in the original VM~\cite{VM}. However, the species-dependent interaction leads to two distinct types of ordering as exemplified in Fig.~\ref{fig1}(b) and (c). 

\begin{figure*}[t]
\centering
\includegraphics[width=14cm]{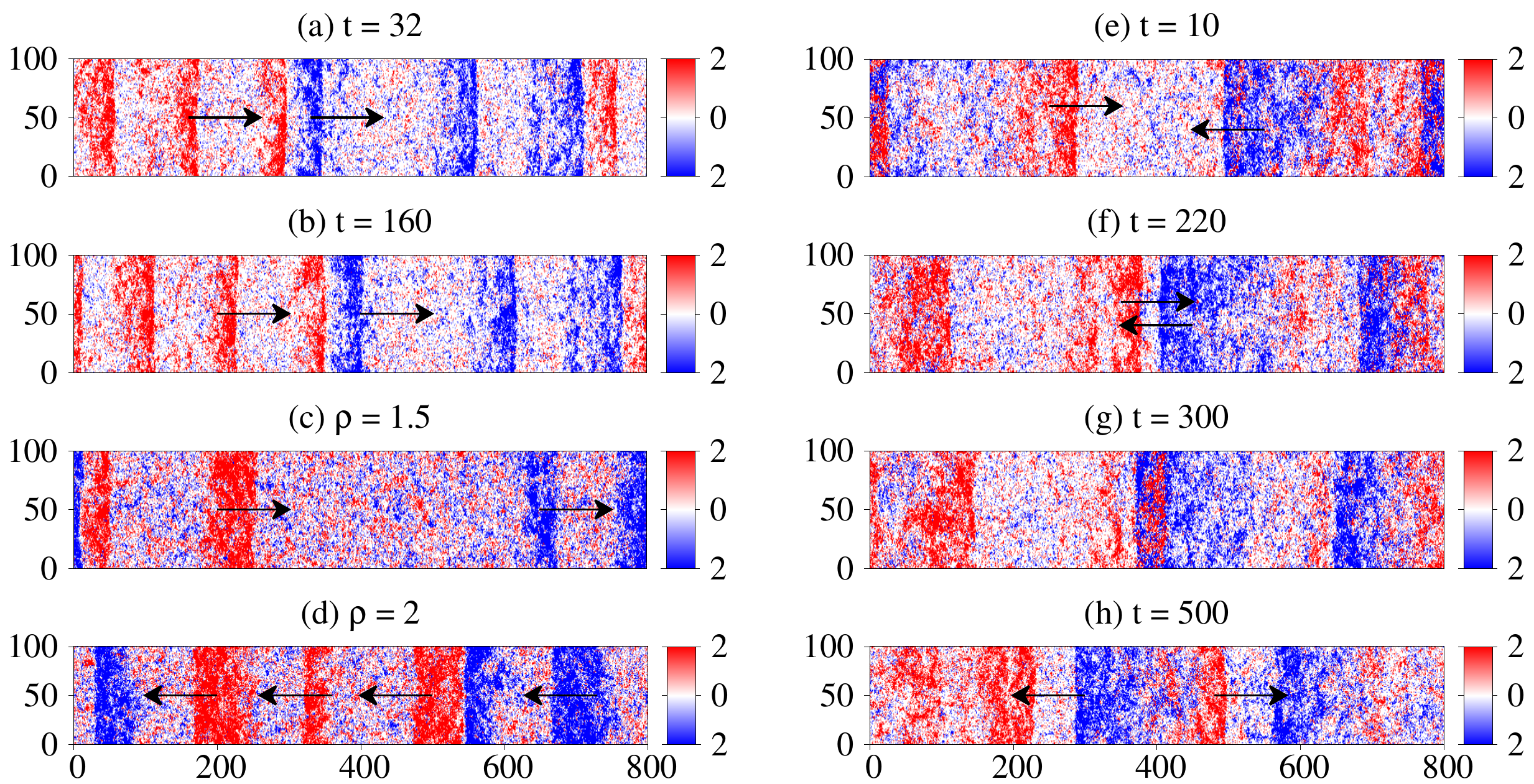}
\caption{(color online) Snapshots of the time evolution of the TSVM ($L_x=800$, $L_y=100$, $v_0=0.5$). A~(B)-particles are represented with red~(blue) dots, and a local particle density is color coded according to the color bar. (a--b) Time evolution of a PF state~($\eta=0.3$, $\rho=1$). Bands of A and B species move in the same direction indicated by the arrows. (c--d) Comparison of a PF state at two different densities~($\eta=0.4$). There are two bands for $\rho=1.5$ and three bands for $\rho=2$. (e--h) Time evolution of an APF state~($\eta=0.3, \rho=1.4$). The two central bands approach each other ($t=10$ and $t=220$), collide ($t=300$), and pass by ($t=500$), as indicated by the arrows.}
\label{fig2}
\end{figure*}

Hereafter, we consider a rectangular geometry of large aspect ratio $L_x/L_y=8$, if not stated otherwise, to force putative bands to move in either $+x$ or $-x$ direction. Fig.~\ref{fig2} shows detailed snapshots of the time evolution of the TSVM in the phase coexistence region. Each species is microphase-separated forming traveling bands, A-bands and B-bands, and there are two types of dynamic states: (i) A- and B-bands move in the same direction, which we will denote as a ``parallel flocking''~(PF) and (ii) A- and B-bands move in the opposite direction, which we will denote as an ``anti-parallel flocking''~(APF). We will investigate the dynamical properties of the PF and APF states to understand the global phase diagram of the system.

\begin{figure}[t]
\centering
\includegraphics[width=\columnwidth]{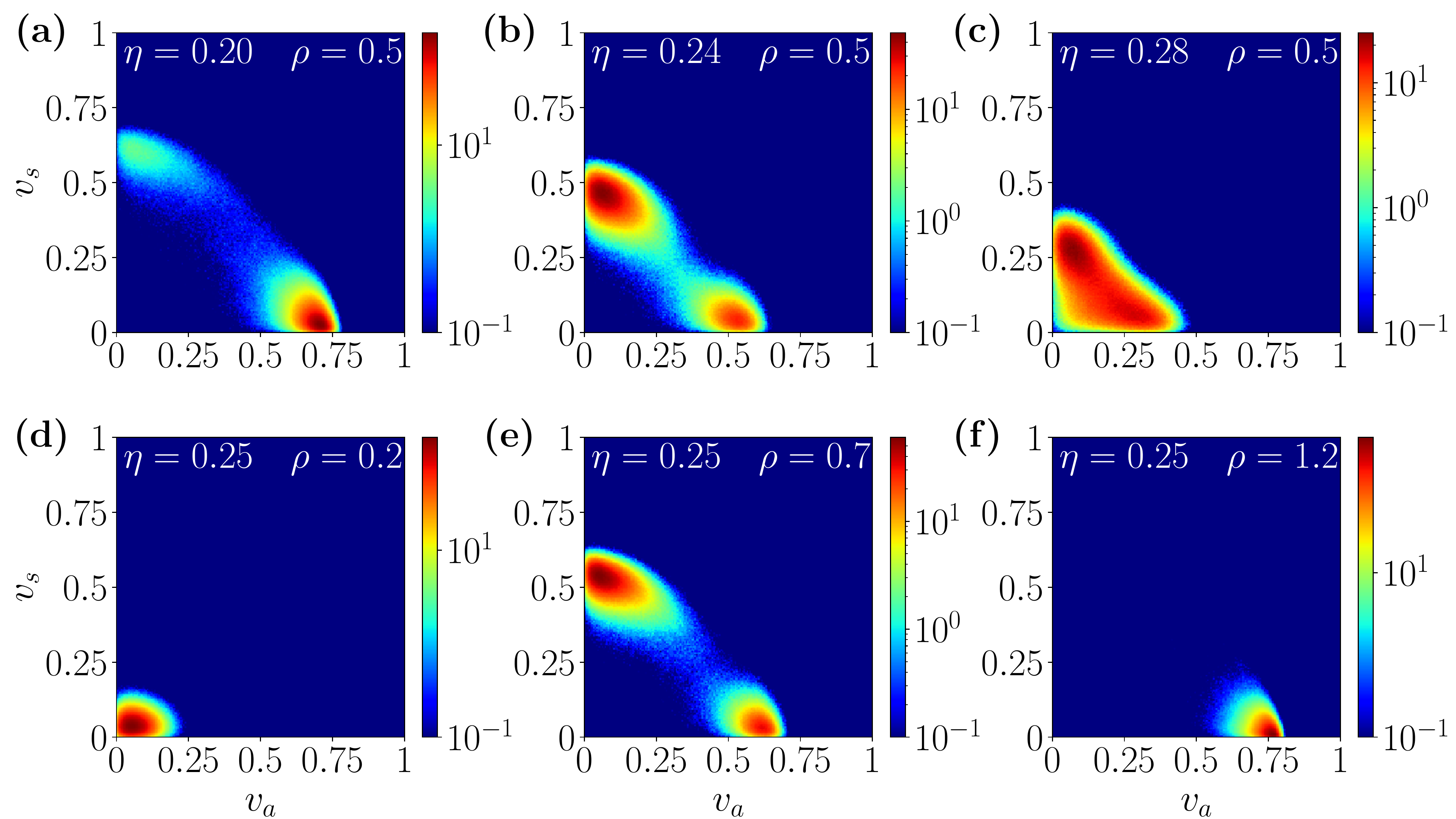}\\
\caption{(color online) Probability distribution $P(v_s, v_a)$ at fixed
$\rho=0.5$ with varying $\eta$~(top row) and at fixed $\eta=0.25$ with
varying $\rho$~(bottom row). The velocity modulus is fixed to $v_0 = 0.5$.
A single peak in (d) signifies the disordered
gas phase while two peaks in (a), (b), (c) and (e) indicates stochastic
switching between the PF and APF states. The ordered liquid phase is
described in (f), which is characterized with a single peak 
with finite APF order parameter.}
\label{fig3}
\end{figure}

\begin{figure}[t]
\centering
\includegraphics[width=\columnwidth]{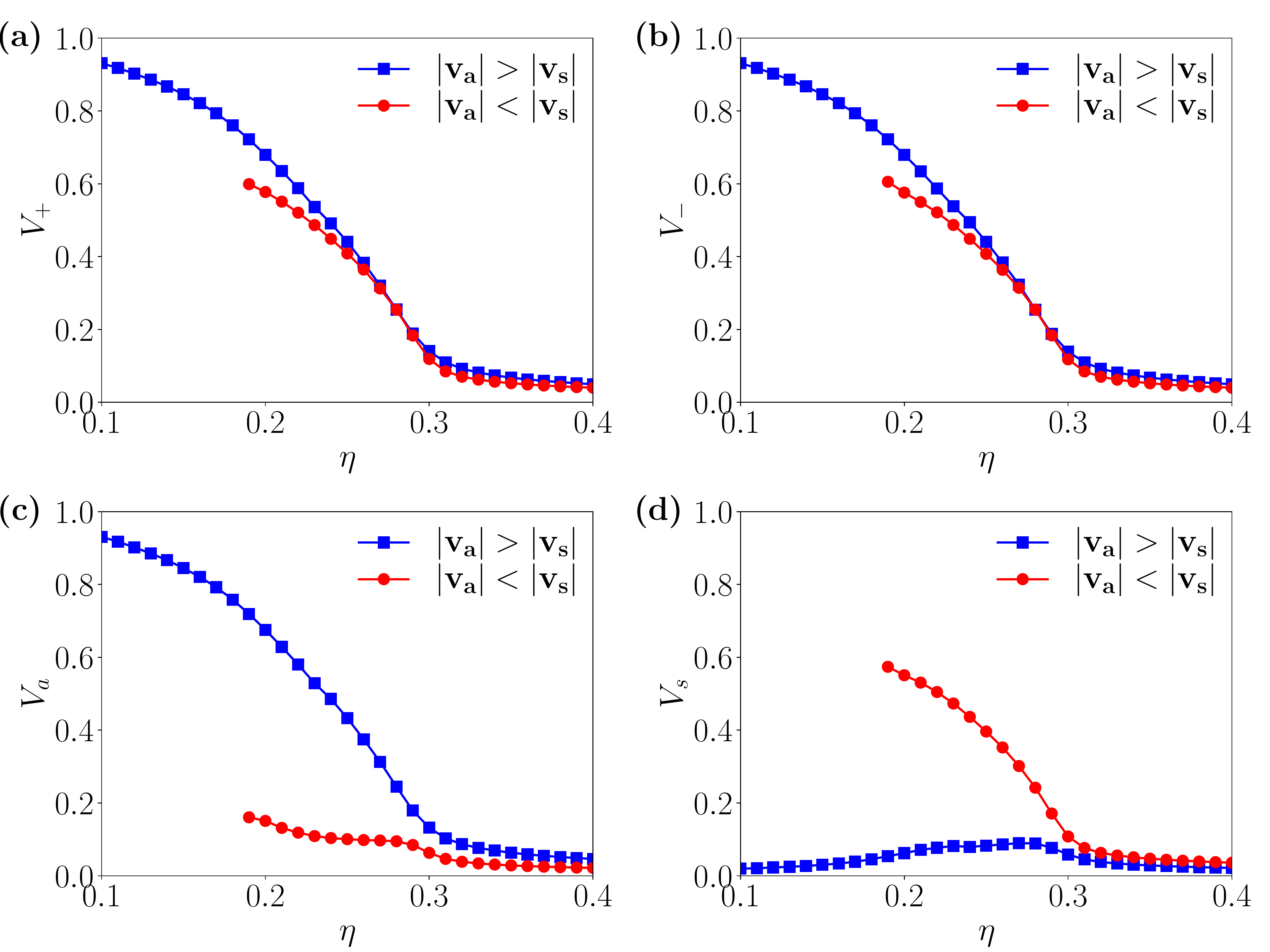}\\
\caption{(color online) Order parameters in the restricted 
    APF~(square symbols) and PF~(circular symbols) ensembles versus 
$\eta$ for $v_0 = 0.5$, $\rho=0.5$, $L_x = 200$, and $L_y = 25$.}
\label{fig4}
\end{figure}

{\bfseries Order parameter} - 
For a quantitative analysis, we introduce an order parameter for the
collective motion. The instantaneous order parameters for the collective
motion of the A and B species are given by 
\begin{equation}
\begin{aligned}
{\bm v_+}(t)&=\frac{1}{N_A} \sum_{i \in A} \bm{\sigma}_i^t, \\
{\bm v_-}(t)&=\frac{1}{N_B} \sum_{i \in B} \bm{\sigma}_i^t \, .
\end{aligned}
\end{equation}
The flocking order parameters are defined as $V_{\pm} =
\langle{|\bm{v}_{\pm}(t)|}\rangle$ where $\langle{(\dots)}\rangle$ 
denotes the time average in the steady state and the ensemble average over
independent runs. 
These order parameters should be the same~($V_+ = \ V_-$) and become nonzero
when the collective motion sets in.
The PF and APF states are distinguished with 
\begin{equation}
\label{OP_Eq}
\begin{aligned}
    {\bm v_s}(t) &= \frac{1}{N}\sum_{i=1}^N \bm{\sigma}_i^t = \bm{v}_+(t) +
    \bm{v}_-(t), \\
    {\bm v_a}(t) &= \frac{1}{N}\sum_{i=1}^N s_i^t \bm{\sigma}_i^t =
    \bm{v}_+(t) - \bm{v}_-(t) \, ,
\end{aligned}
\end{equation}
from which we define the order parameters $V_s = \langle |\bm{v}_s(t)|\rangle$
for the PF state and $V_a = \langle | \bm{v}_a(t) |\rangle$ for the APF state.
We expect that $V_s > 0$ and $V_a=0$ in the PF state while $V_s = 0$ and
$V_a > 0$ in the APF state in the thermodynamic limit.

The probability distribution $P(v_a, v_s)$ constructed from the steady state 
time series of $v_s = |\bm{v}_s|$ and $v_a = |\bm{v}_a|$ from independent
runs is presented in Fig.~\ref{fig3}. When the noise is large or the density
is small, the probability distribution has a single peak near $v_a = v_s
=0$, which represents the disordered gas phase. Interestingly, the
probability distribution in the intermediate parameter regime has two peaks,
which manifests the existence of the PF and APF states. 
The two peaks structure also indicates stochastic switches between the two dynamic states in the steady state. The switch dynamics will be studied 
in detail below.

The order parameter time series may include the contributions from 
both dynamic states. In order to characterize the PF and APF states separately, we measure the order parameters $V_{\pm}$ and $V_{s,a}$ using a restricted ensemble
average. The system is assigned to be in a PF ensemble when $|\bm{v}_s(t)| > |\bm{v}_a(t)|$ or in an APF ensemble otherwise. Order parameters averaged within the restricted ensemble are plotted in Fig.~\ref{fig4}. These plots show that collective motion sets in below a certain noise strength and above a certain density. They also show that the APF order is stronger than the PF order in the sense that the order parameter in the APF ensemble takes a larger value than that in the PF ensemble. $V_a$ in the PF ensemble is larger than $V_s$ in the APF ensemble, which indicates that fluctuations are stronger in the PF state. The PF ensemble data are missing for $\eta \lesssim  0.2$, which will be addressed later.

The APF order is stronger than the PF order since the ordering can be
enhanced by exploiting the inter-species anti-alignment interactions.
In terms of a variable $\bm{\alpha}_i \equiv s_i \bm{\sigma}_i$, the
alignment rule in Eq.~\eqref{sigma2} can be rewritten as
\begin{equation}
\bar{\bm{\alpha}}_i^t = \sum_{j\in \mathcal{N}_i} \bm{\alpha}_j^t.
\end{equation}
Namely, each particle aligns its
$\bm{\alpha}$ variable with those of its neighboring particles 
regardless of the particle species. This representation
demonstrates that the PF state is stable only when condensates of different
species, having opposite $\bm{\alpha}$ vectors, are spatially 
separated~(see Fig.~\ref{fig2}). One the other hand, in the APF state, condensates of different species, having parallel $\bm{\alpha}$ vectors, are not mutually exclusive~(see Fig.~\ref{fig2}). Thus, particles in the APF state 
see more ``correctly aligned" neighbors (from its own and the other species), 
which decreases fluctuations.

The PF and APF ordering reveals the mechanism to achieve flocking in a two species population with frustrating interactions: the two species may be separated spatially to avoid the anti-alignment interaction and move in the same direction~(PF), or two species may move in the opposite direction to satisfy the anti-alignment interaction~(APF).

{\bfseries Moving bands} - As seen in Fig.~\ref{fig2}, particles in the coexistence regime are organized into an array of randomly spaced ordered bands propagating in the $+x$ or $-x$ direction and spanning the system along the $y$ direction. This arrangement of finite-width bands is known as \emph{microphase separation}~\cite{SolonVM,acm}, which differs from the conventional liquid-gas phase separation observed in the flocking model with discrete symmetry~\cite{acm,aim,apm}. 

\begin{figure}[t]
\centering
\includegraphics[width=\columnwidth]{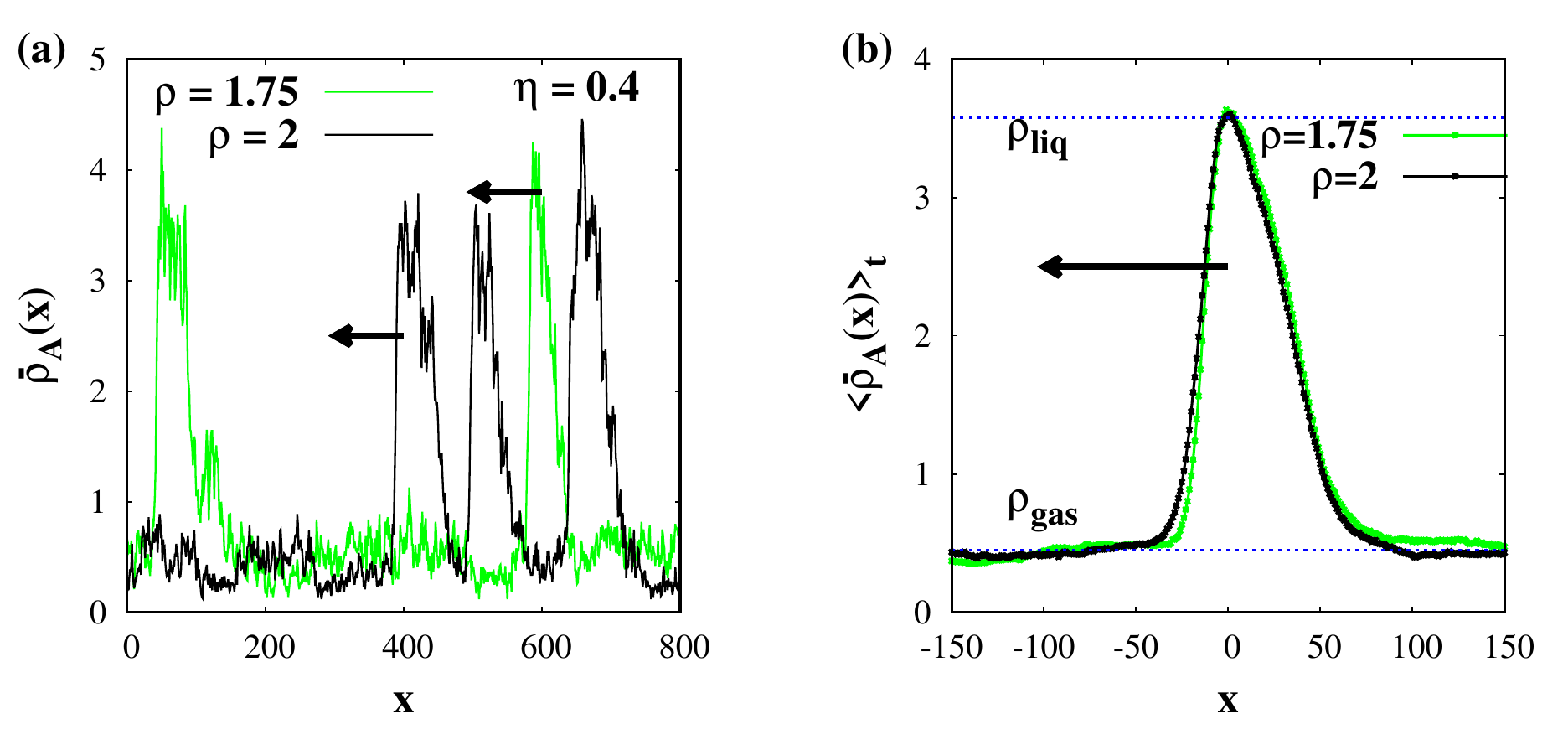}\\
\caption{(color online) Phase-separated (a) instantaneous and (b)
time-averaged  density profiles of species A for two values of $\rho$ with
fixed $\eta = 0.4$ and $v_0=0.5$. Black arrows indicate the direction
of propagation.}
\label{fig5}
\end{figure}

The bands appear as a density wave in the time-dependent density profiles
$\bar\rho_{A,B}(x,t)$ as shown in Fig.~\ref{fig5}(a), where the overbar 
refers to an average along the $y$-direction.
The stationary average shape is obtained from a running average of the 
time-dependent profile:
\begin{equation}
\langle \bar{\rho}_A(x) \rangle_t = \frac{1}{n_p} \sum_{k=1}^{n_p}
\bar{\rho}_A(x+x_k,t_k)\, ,
\end{equation}
where $n_p$ is the number of instantaneous profiles and $x_k$ denotes
the peak position at time $t_k$. Fig.~\ref{fig5}(b) shows that
the density wave has an asymmetric shape signifying its propagating
direction. The density wave moves on a uniform background, whose density 
will be denoted as $\rho_{\rm gas}$. 
The peak density will be denoted as $\rho_{\rm liq}$.

The average shape of the band helps us decipher the global phase diagram. For given value of $\eta$ and $v_0$ the average shape of the band shown in Fig.~\ref{fig5}(b) does not change as one varies the overall density. Instead only the number of bands increases~[see Figs.~\ref{fig2}(c--d)]. It indicates that 
$\rho_{\rm gas}$ and $\rho_{\rm liq}$ are the binodal densities separating the two homogeneous phases, liquid (at low noise and high density) and gas~(at high noise and low density), from the microphase-separated coexistence phase.

\begin{figure}[t]
\centering
\includegraphics[width=6cm]{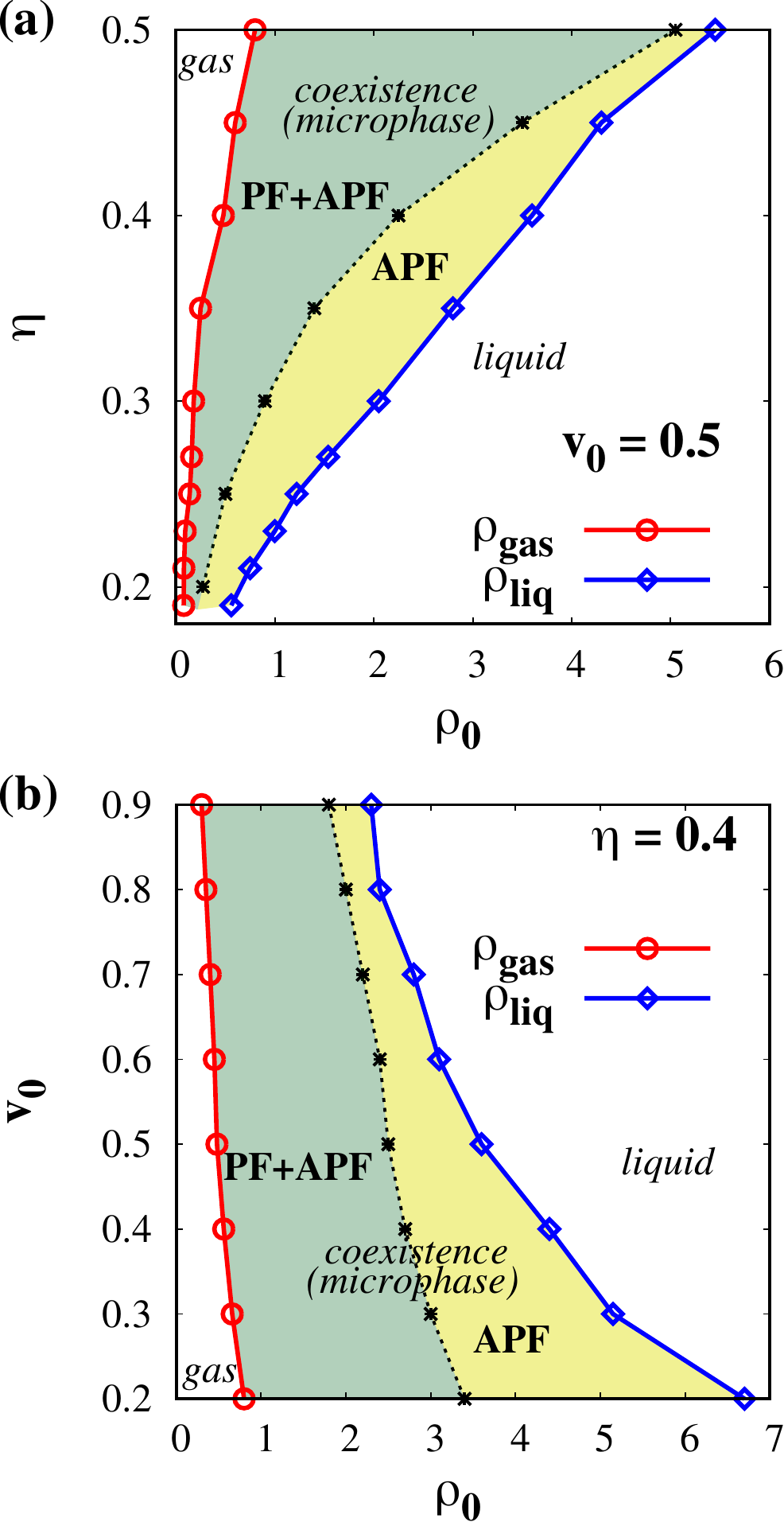}\\
\caption{(color online) (a) $\eta-\rho_0$ phase diagram for fixed $v_0=0.5$
and (b) $v_0-\rho_0$ phase diagram for fixed $\eta=0.4$ of the TSVM.
The binodals $\rho_{\rm liq}$ and $\rho_{\rm gas}$ are the boundaries of the coexistence phase in the thermodynamic limit, obtained by extrapolating the finite size boundaries to an infinite system size. The black dotted lines in the coexistence demarcates the boundary for the PF state and is also independent of system size: it indicates the boundary between the region of the coexistence phase where one observes PF-APF transition for any finite system size and the region where one does not.}
\label{fig6}
\end{figure}

{\bfseries Phase diagram} - We summarize our findings with the noise-density 
($\eta-\rho_0$) and speed-density ($v_0-\rho_0$) phase diagrams in Fig.~\ref{fig6}. Note that $\rho_0 = \rho/2$ is the density of either A or B species. In the gas phase, particles are distributed uniformly and the flocking order parameters vanish.

In the coexistence phase~($\rho_{\rm gas}\leqslant \rho_0 \leqslant \rho_{\rm liq})$, 
the area fraction $\phi$ of the liquid bands of each species satisfies the relation
$\rho_{\rm gas} (1-\phi)+  c \rho_{\rm liq}\phi = \rho_0$, where $c\rho_{\rm liq}$ is the average density of a band with a positive constant $c\leqslant 1$. It leads to 
\begin{equation}
\phi = \frac{\rho_0 - \rho_{\rm gas}}{c \rho_{\rm liq} - \rho_{\rm gas}}.
\end{equation}
Since the coexistence phase bands are not perfectly rectangular, we introduce an $O(1)$ parameter $c$ (unknown but measurable) to express the total area of a band. $c=1$ signifies an ideal rectangular band but due to the fluctuation and off-lattice geometry, obtaining a perfectly rectangular band is impossible and therefore, $c<1$ but close to 1.

The A-bands and B-bands repel each other in the PF state. If the repulsion is perfect, the PF state is constrained by the condition $2 \phi \leqslant 1$, or equivalently,
\begin{equation}
\label{PF_limit}
\rho_{\rm gas} \leqslant \rho_0 \leqslant \frac{1}{2}\left(\rho_{\rm
gas}+ c \rho_{\rm liq}\right)\, .
\end{equation}
From Eq.~\eqref{PF_limit}, the boundary between the regions where the PF state still exists and where it does not is somewhere between $\rho_{\rm gas}$ and $\rho_{\rm liq}$. For $c=1$, the boundary is approximately in the middle of $\rho_{\rm gas}$ and $\rho_{\rm liq}$, whereas for $c < 1$, it is slightly shifted to the left towards $\rho_{\rm gas}$. In contrast, the micro-separated APF state can be observed in the entire coexistence phase with $\rho_{\rm gas} \leqslant \rho_0 \leqslant \rho_{\rm
liq}$. Therefore, the coexistence region is further separated into the 
PF+APF region where the system stochastically switches between the
two states and the APF region where only the APF state is
stable. The boundary between the two regions is drawn with the dotted 
line in the phase diagram. 
In Fig.~\ref{fig4}, we have already observed that the PF state is not stable
when the noise strength is low enough. Numerically, the boundary 
is obtained by estimating the density beyond which the PF ensemble is
absent.
Note that the PF state is observed beyond the
approximate limit in Eq.~\eqref{PF_limit}. Nevertheless, the mutual
repulsion between A- and B-bands successfully explains that the PF state is
stable only in the low density part of the coexistence region. 

In the liquid phase, the continuous orientational symmetry is spontaneously 
broken and the system exhibits a long range orientational order~(see
Fig.~\ref{fig11}(a) of Appendix~\ref{appNF}). As mentioned earlier only the 
APF state is stable in the liquid phase. The gas phase and the liquid 
phase have a different symmetry for which reason the two binodals cannot 
merge in a single critical point as in discrete flocking models \cite{aim,apm,acm}. 
The liquid phase is further characterized by
giant number fluctuations~(see Fig.~\ref{fig11}(b) of Appendix~\ref{appNF}) as
in the original Vicsek model~\cite{SolonVM}. As conjectured in Ref.~\cite{SolonVM}, the giant number fluctuations in the liquid phase 
are believed to be responsible for the instability of a single 
macroscopic liquid cluster in the coexistence region leading to microphase separation instead~\cite{SolonVM}. 

\section{PF-APF transitions}
\label{s4} 

\begin{figure}[t]
\centering
\includegraphics[width=\columnwidth]{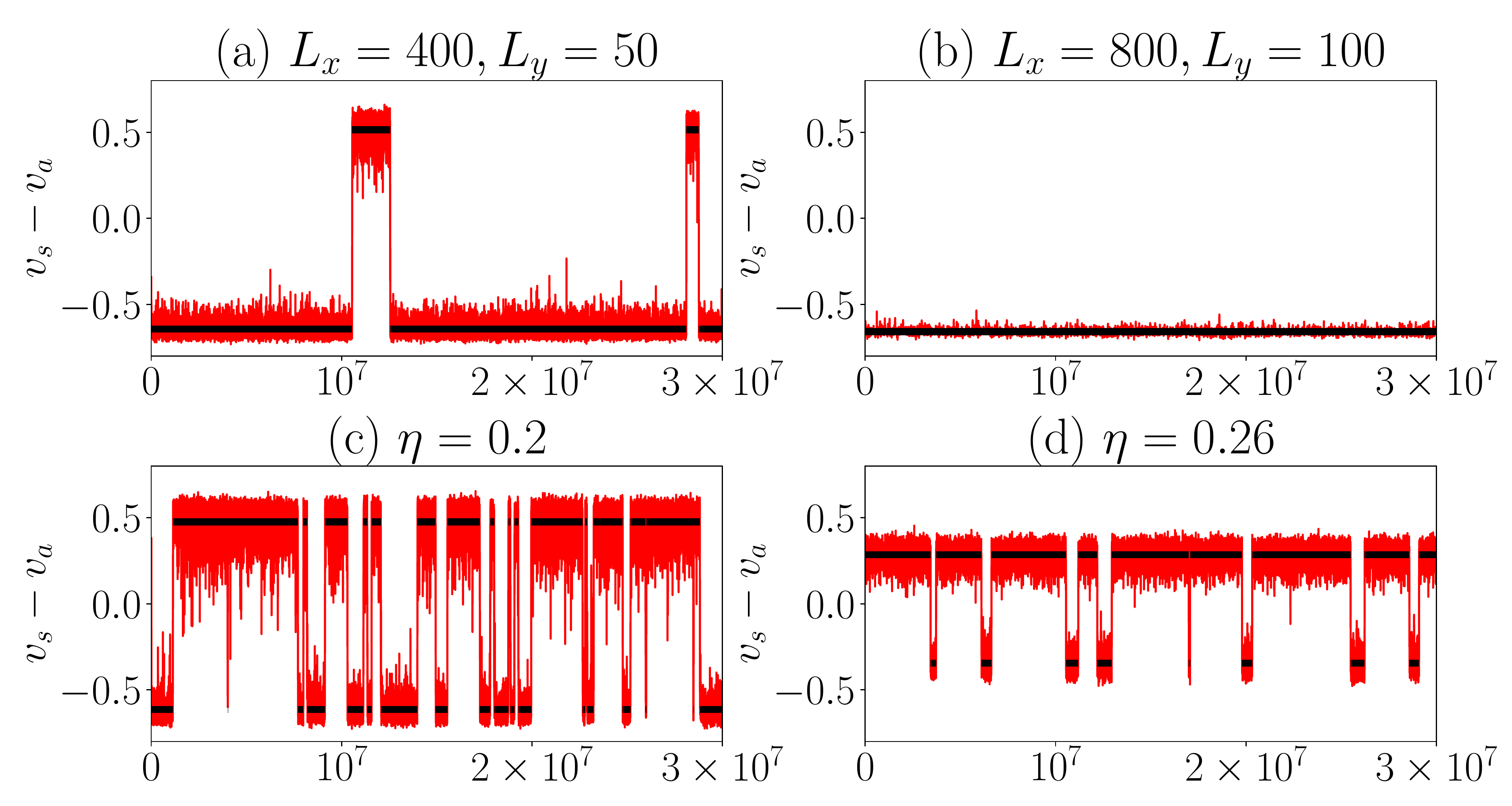}\\
\caption{(color online) Comparison of the time series of $v_s-v_a$ 
at two system sizes (with $\eta=0.2$) in (a,b), and at two noise strengths
(with $L_x=512$, $L_y=32$) in (c,d). The thick solid line segments 
represent the time intervals during which the system dwells in the
PF or APF states. $v_0=0.5$ and $\rho=0.5$.}
\label{fig7}
\end{figure}
In the PF+APF region, the system switches back and forth between PF and 
APF states. The time series of $(v_s - v_a)$ shown in Fig.~\ref{fig7}
demonstrates the stochastic transitions. We characterize the stochastic transitions with the dwell time distribution and observe that the dwell time distribution has an exponential tail with a characteristic time almost equal to the mean dwell time.
The quantity $(v_s-v_a)$ fluctuates around a positive value $p$ in the PF state and a negative value $-q$ in the APF state. We identify an APF-to-PF transition by the moment when $(v_s-v_a)$ exceeds a threshold value $r p$, and a PF-to-APF transition by the moment when $(v_s-v_a)$ falls below $-r q$ with a constant $0< r<1$. 
Then, the time series leads to a sequence of alternating dwell times $\cdots
\to t_{\rm PF} \to t_{\rm APF} \to t_{\rm PF}' \to t_{\rm APF}' \to \cdots$.
It is useful to introduce the thresholds with a positive constant $r$ since for $r=0$, microscopic fluctuations during a transition would be regarded as multiple transitions whose time scale is much shorter than the macroscopic dwell time. Here we choose $r=0.7$ and demonstrate the obtained transition sequences between the PF and APF states in Fig.~\ref{fig7}.

\begin{figure}[t]
\centering
\includegraphics[width=\columnwidth]{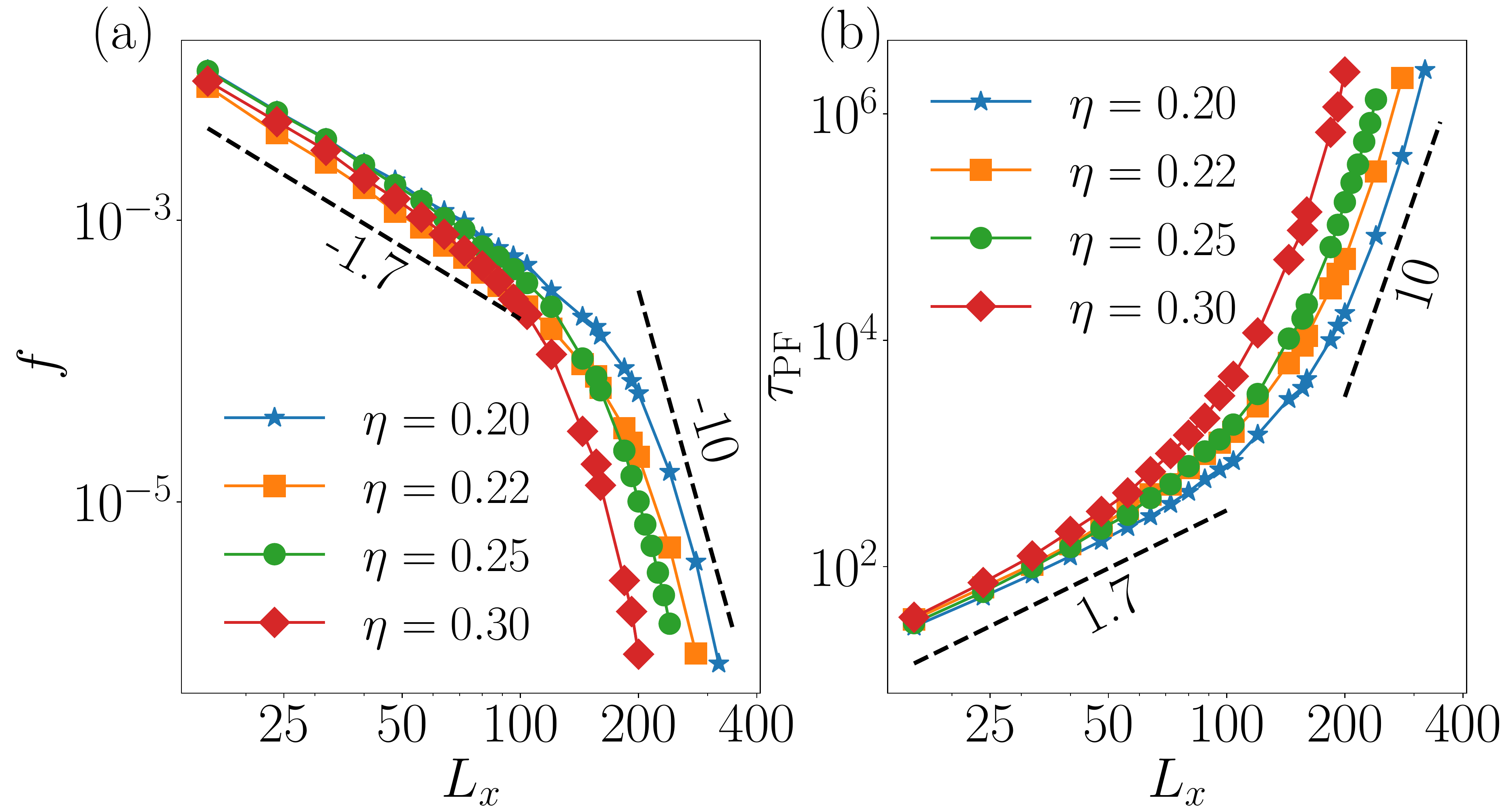}\\
\caption{(color online) (a) Average frequency $f$ of transitions between the
APF and PF states and (b) average time $\tau_{\rm PF}$ spent by the system
in the PF state as a function of linear system size $L_x$ (on a log-log
scale), where $L_y=L_x/8$. The dashed lines with different slopes indicate a crossover in the
scaling law at a crossover length $L_{x,c}$ depending of $\eta$ and $\rho$. 
Parameters: $v_0=0.5$, $\rho=0.35$ ($\eta=0.2$), $\rho=0.5$ ($\eta=0.22$), $\rho=0.6$ ($\eta=0.25$), and $\rho=1$ ($\eta=0.3$).}
\label{fig8}
\end{figure}

Any transition requires the velocity reversal of all bands of one species,
which is expected to take longer as the system size increases.
Fig.~\ref{fig8} confirms this expectation by analyzing the finite-size dependence of the transition frequency $f$ and the PF state dwell time 
$\tau_{\rm PF}$ by varying $L_x$ with fixed aspect ratio
$L_x/L_y=8$. The transition frequency $f$ is defined as the number of transitions per unit time and the dwell times ($\tau_{\rm PF}$ or $\tau_{\rm APF}$) are the average time spent by the system in either of the states. After the system reaches the steady state, the number of PF-to-APF and APF-to-PF transitions are identified and recorded for a long time ($t_m \sim 10^7$) using the method mentioned above. $f$ is then computed by dividing the total number of transitions by $t_m$ and the ratio of the total time the system spent in the PF (APF) state and the number of PF-to-APF transitions (APF-to-PF transitions) produce $\tau_{\rm PF}$ ($\tau_{\rm APF}$).
Interestingly, both the transition frequency and the dwell time exhibit a
sharp crossover at a crossover length scale $L_{x,c}$: for $L_x \ll L_{x,c}$,
the average dwell time increases algebraically as 
$\tau_{\rm PF} \propto L_x^{z_1}$ with $z_1 \simeq 1.7$. 
For $L_x \gg L_{x,c}$ it increases much faster as $\tau_{\rm PF} \sim
L_x^{z_2}$ with $z_2 \simeq 10.0$, which is so large 
that we cannot exclude an exponential scaling.

\begin{figure}[t]
\includegraphics*[width=\columnwidth]{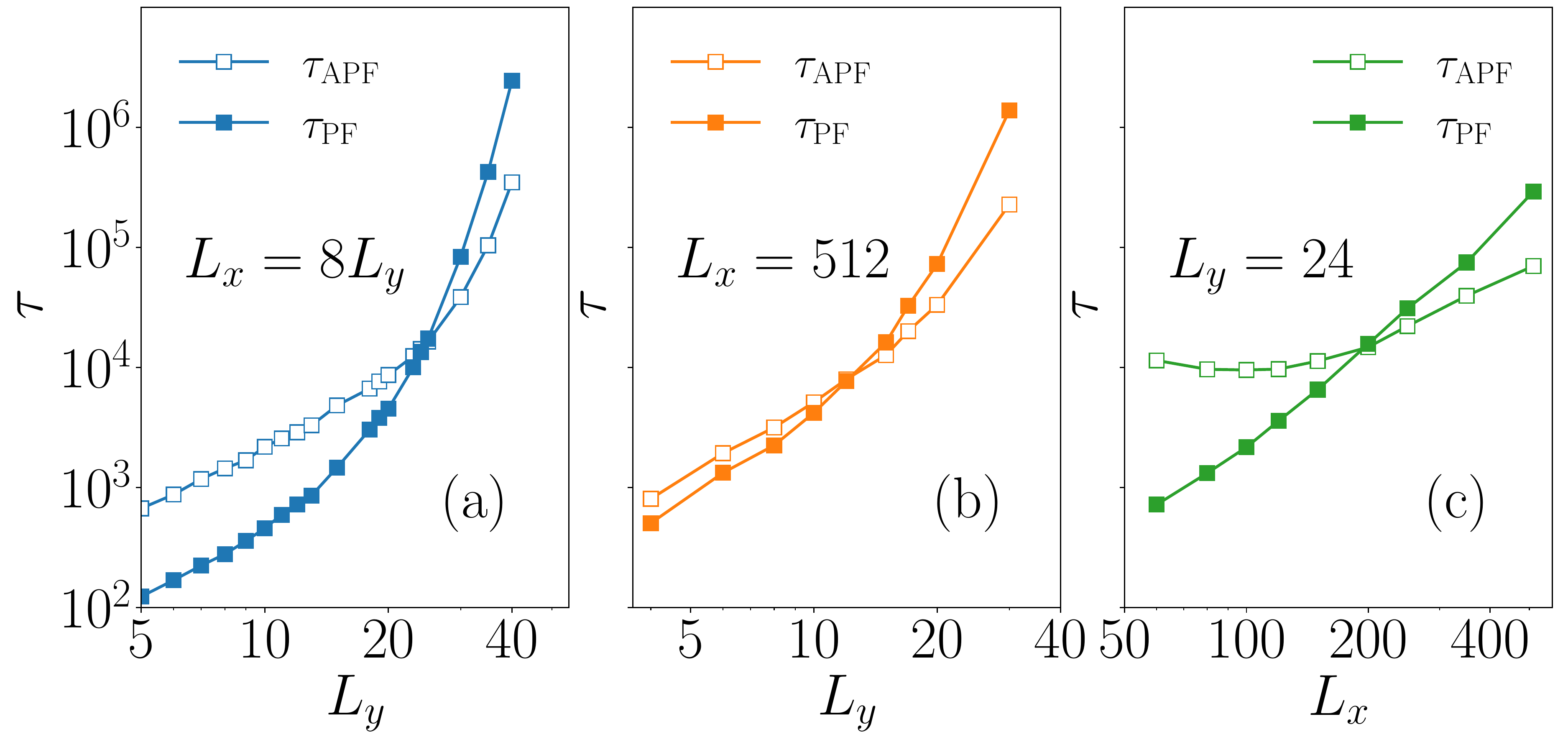}\\
\caption{System size dependence of the average dwell time of the APF state~(empty
symbols) and the PF state~(filled symbols) with fixed aspect ratio
$L_x/L_y=8$ in (a), fixed width $L_x=512$ in (b), and fixed height $L_y=24$ in (c). 
All measurements are done with $v_0=0.5$, $\rho=0.35$, and $\eta=0.2$.}
\label{fig9}
\end{figure}

To shed light on the origin of the observed crossover we studied the 
dependence of the dwell time as a function of the system height $L_y$.
Fig.~\ref{fig9}(a) compares the dwell times of the PF and APF states with
fixed $L_x/L_y = 8$. Both quantities show the crossover at the same
length scale. We also determined the dwell times for a large but fixed 
value of $L_x=512$ as a function of $L_y$. Fig.~\ref{fig9}(b) 
demonstrates that a similar crossover occurs at a height $L_{y,c} \simeq 10$. 
On the other hand, the dwell time does not have a crossover as $L_x$ is varied with fixed $L_y = 24$~[see Fig.~\ref{fig9}(c)].

\begin{figure}[t]
\centering
\includegraphics[width=\columnwidth]{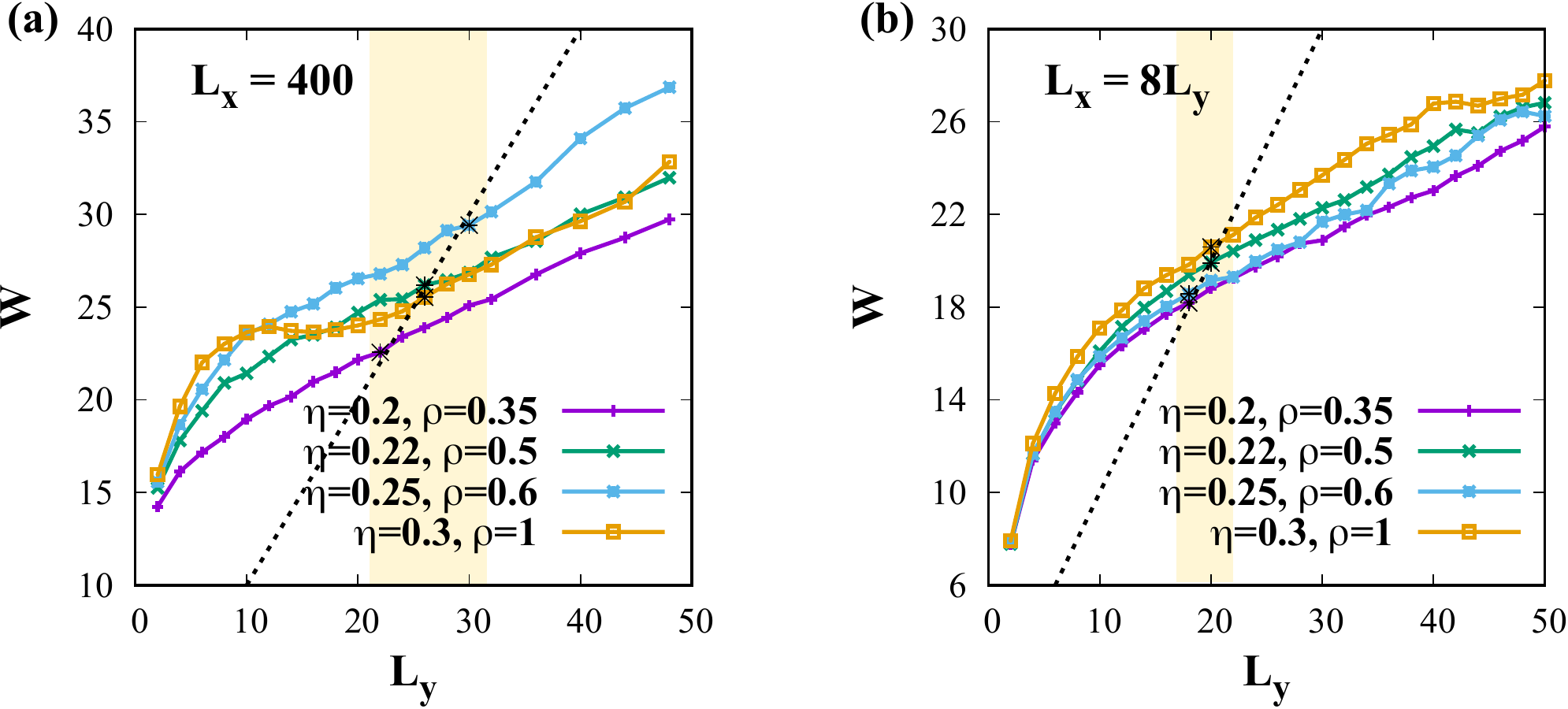}\\
\caption{(color online) Band width $W$ versus $L_y$ for various $(\eta,\rho)$ combinations. In (a) $L_x$ is fixed to 400, and in (b) it is $L_x=8L_y$.
In (a), the points (black cross) where $W\approx L_y$ [for each set of $(\eta,\rho)$] are for $20<L_y<32$ whereas in (b) the $W\approx L_y$ points are around $L_y \approx 20$. The black dashed line represents $W=L_y$ and the shaded region is a guide to the eyes.
The width is computed from the time- and
ensemble-averaged density profile of bands and
is defined as the distance between the two points in the density
profile (c.f. Fig.~\ref{fig5}b) where the density is equal to 
$(\rho_{\rm gas}+\rho_{\rm liq})/2$.}
\label{fig10}
\end{figure}

The width $W$ of the bands also depends on the system height $L_y$,
as shown in Fig.~\ref{fig10}, and is larger than $L_y$ for small $L_y$
and smaller for large $L_y$. Remarkably, the value of $L_y$ where it
equals $W$ agrees roughly with the value of $L_y$, where the crossover
in dwell time occurs. These results coherently suggest that the crossover
occurs when the system height~$L_y$ is comparable to the band width~$W$, which appears plausible due to the following reason:  

We observe that in PF-to-APF transition a single band can spontaneously reverse its direction of motion, by first dissolving via fluctuations and then rebuilding with opposite velocity. The subsequent inevitable collision with the other bands of the same species then reverses them, too. A fluctuation induced transition of a whole band from one metastable configuration (e.g. right moving) to another (then left moving) is a rare event whose probability decreases with the size of the band, i.e. with increasing $L_y$, as can be seen in Fig.~\ref{fig8} and Fig.~\ref{fig9}, but faster for $L_y>W$ than for $L_y<W$. We think that this is due to a change in the characteristics of the necessary fluctuation reverting a band: for $L_y<W$, these fluctuations have to be predominantly correlated in the longitudinal direction, i.e. in the direction of motion of the band, whereas for $L_y>W$, they have to be correlated in the transverse direction, i.e. perpendicular to the direction of the band motion (see Appendix~\ref{AppB} for a demonstration). Probably one could quantify this picture by a detailed study of the density-density correlation functions in x and y direction inside the band, which we leave for future investigations. We also want to point out that for $L_y \ll W$, the band looks like a piston moving “longitudinally” within a pipe, but this longitudinal movement is different from the formation of longitudinal bands or “lanes” observed in other flocking models ~\cite{apm, Ginelli2010}.

\section{Discussion}
\label{s5}
To summarize, we have shown that the flocking transition in the two-species Vicsek model (TSVM) is in many aspects analogous to the original Vicsek model:
it has a liquid-gas phase transition and displays micro-phase separation in the coexistence region where multiple dense liquid bands propagate in a gaseous background. The interesting feature of the TSVM is the appearance of two dynamical states in the coexistence region: the PF (parallel flocking) state in which all 
bands of the two species propagate in the same direction, and the APF (anti-parallel flocking) state in which the bands of species A and species B move in opposite directions. 

Due to the anti-alignment rule between different species,
A and B bands (or clusters) moving in opposite directions do not disturb each other upon collision, on the contrary they even stabilize each other. This is markedly different in the PF state: here the ant-alignment rule destabilizes the bands (or clusters) 
of different species moving in the same direction upon contact, for which reason they are only stable when they move in some distance to each other in the same direction. Consequently,
PF states only occur in the low-density part of the coexistence region -
at higher densities and in particular in the liquid phase, only the APF state occurs.

When PF and APF states exist in the low-density part of the 
coexistence region they perform stochastic transitions from one to the other.
Their frequency decreases with increasing system size as
the dwell times in the two states increase. The system size 
dependence shows a crossover from a power law with a dynamical 
exponent $z_1$ to a much steeper power law with a much larger 
dynamical exponent $z_2$ (or exponential dependence). 
The crossover is related to a change in the nature of the 
fluctuations when the system size in $y$-direction increases 
beyond the width of the bands moving in $x$-direction.

Here we presented only results for the basic version of the TSVM,
but it would be interesting to study some variations such as the TSVM with different species densities ($N_A \neq N_B$) or different species speeds ($v_0^A$ $\neq v_0^B$). 
One could also consider the TSVM with spatial heterogeneity where in one region, $v_0^A$ $> v_0^B$ whereas $v_0^B$ $> v_0^A$ in the other region (c.f. Ref.
\cite{activitylandscape} and references therein). Preliminary investigations of these variants of the TSVM show interesting collective dynamics such as increasing the number density of one species ($N_A \neq N_B$, $N_A+N_B=N$) destroys the APF state and the system converts to the original VM and a change in band formation due to different species velocities in different region. Another interesting prospect would be to investigate the multi-species effect on the well known discrete flocking models, such as the active Ising model~\cite{aim} or the active clock model~\cite{acm}. 

Whereas the alignment rule of the original Vicsek model is directly motivated by the collective behavior of animal flocks, it is hard to think of biological entities that tend to interact via anti-alignment. However, synthetic active matter could be designed to have such interactions. ``Unfriendly" species could for instance be realized by the experimental setup used in Ref. [30], where colloids are activated individually by a laser. The activation strength of each particle was set by a computer that analyzes its current neighborhood. One could also label the particles as A- or B-particle as in our model and instruct the computer for instance to ignore or weight negatively the neighboring B particles when computing the activation strength of an A particle, realizing “unfriendly” species in this way. It would certainly be worthwhile to think about ways to manipulate not only the self-propulsion strength of each particle, but also its direction, and thus realizing alignment or anti-alignment as has been done in Ref.~\cite{fruchart} with programmable robots.

\section{Acknowledgments}
SC, MM and HR were financially supported by the German Research Foundation (DFG) within the Collaborative Research Center SFB 1027. JDN acknowledges the computing resources of Urban Big data and AI Institute (UBAI) at the University of Seoul.


\appendix
\section{Nature of the ordered phase and number fluctuation}
\label{appNF}

\begin{figure}[t]
\centering
\includegraphics[width=\columnwidth]{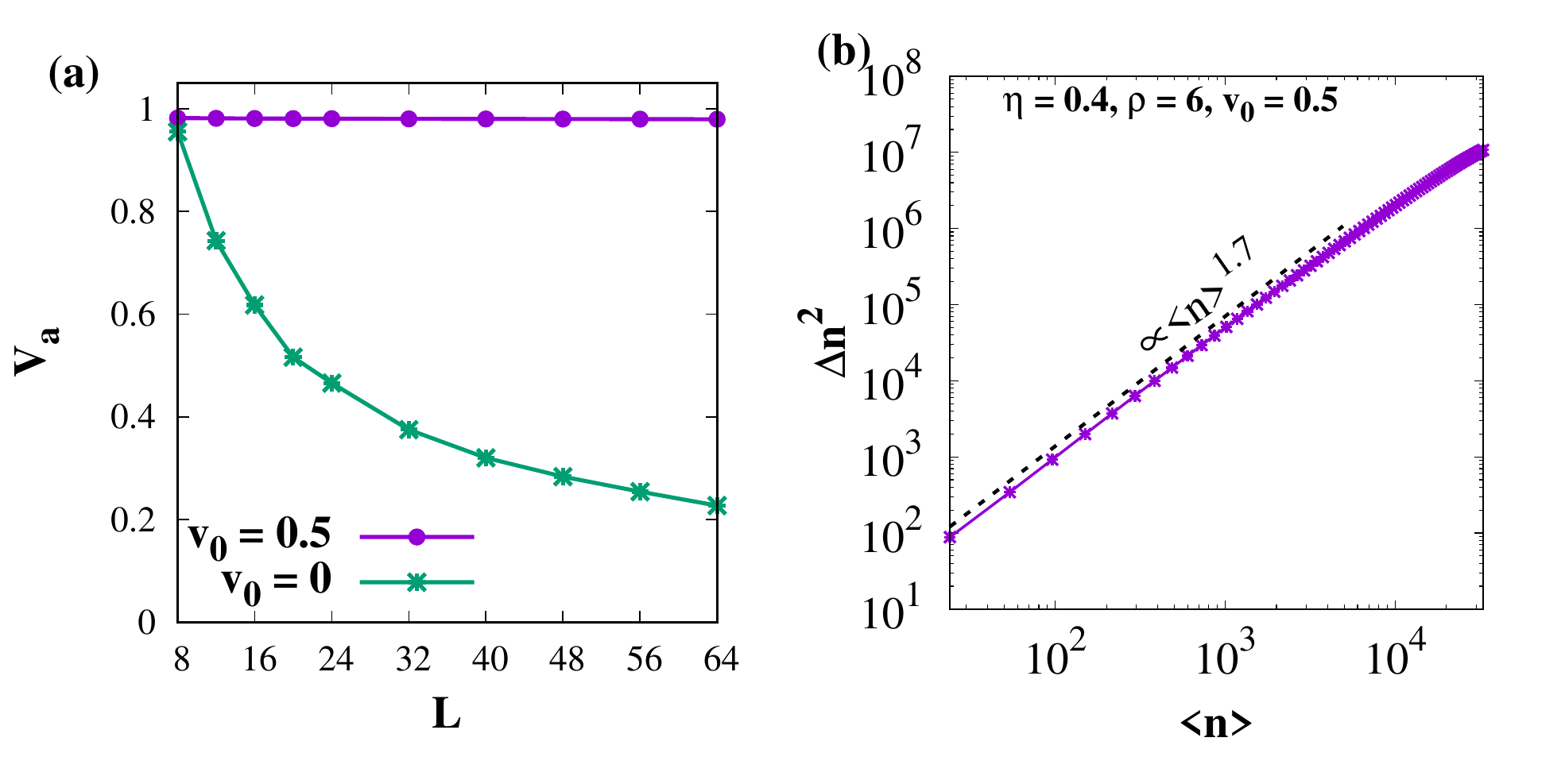}\\
\caption{(color online) (a) $V_a$ against $L$ for the ordered liquid phase: $\eta=0.1$ and $\rho=6$. For $v_0=0.5$, $V_a$ is constant over $L$ whereas for $v_0=0$, we expect $V_a$ decays to zero for $L \to \infty$. (b) Number fluctuations $\Delta n^2=\langle n^2 \rangle-\langle n \rangle^2$ versus average particle number $\langle n \rangle$  in a $200 \times 200$ domain.}
\label{fig11}
\end{figure}

In Vicsek-like models, where particles propel with a constant speed, the ordered state exhibits a true long-range order (LRO) in two dimensions \cite{vicsek97,toner-tu,chate-lro} because the continuous $U(1)$ symmetry is broken spontaneously due to the nonequilibrium activeness of the particles. This spontaneous symmetry breaking is forbidden in equilibrium systems according to the Mermin-Wagner theorem. To understand the nature of ordering of the TSVM liquid phase, we show the  APF order parameter (the liquid phase of the TSVM is an APF state) $V_a$ versus $L$ in Fig.~\ref{fig11}(a) (simulations are done on a square domain of linear length $L$). The data presented is averaged over time and several initial configurations. We note that, $V_a$ remains independent of the system size $L$ for $v_0=0.5$, therefore, one can safely conclude that the system is in the LRO state \cite{shradha,grossman,solon-lro} for the constant-speed version of the model. For $v_0=0$, however, $V_a$ decays to zero 
for $L\to\infty$, probably exponentially since one would expect the TSVM
to behave like a XY spin glass model on a random 2d graph (note that the 
for $v_0=0$ the particles can not move and are frozen in random positions
in 2d, the neighboring particles define the interaction graph and the 
alignment interactions produce a mixture of ferro- and anti-ferromagnetic 
interactions).

Fig.~\ref{fig11}(b) shows the number fluctuation $\Delta n^2=\langle n^2 \rangle-\langle n \rangle^2$ in the liquid phase of the TSVM against the average particle number $\langle n \rangle$ where $n$ is the number of particles in boxes of different sizes $\ell$ included in a $200 \times 200$ domain (for $\ell \leqslant 100$), with $\langle n \rangle = \rho \ell^2$. The fluctuation behaves like $\langle n \rangle ^\xi$ with a fluctuation exponent $\xi \simeq 1.7$ and this value of the fluctuation exponent is close to the exponents extracted for the VM \cite{SolonVM} and the large $q$ limit of the active clock model \cite{acm} and signifies giant fluctuation. This giant number fluctuation is responsible for the microphase in TSVM as hypothesized in Ref.~\cite{SolonVM}.

\section{Longitudinal and transverse density waves}
\label{AppB}
\begin{figure}[t]
\centering
\includegraphics[width=\columnwidth]{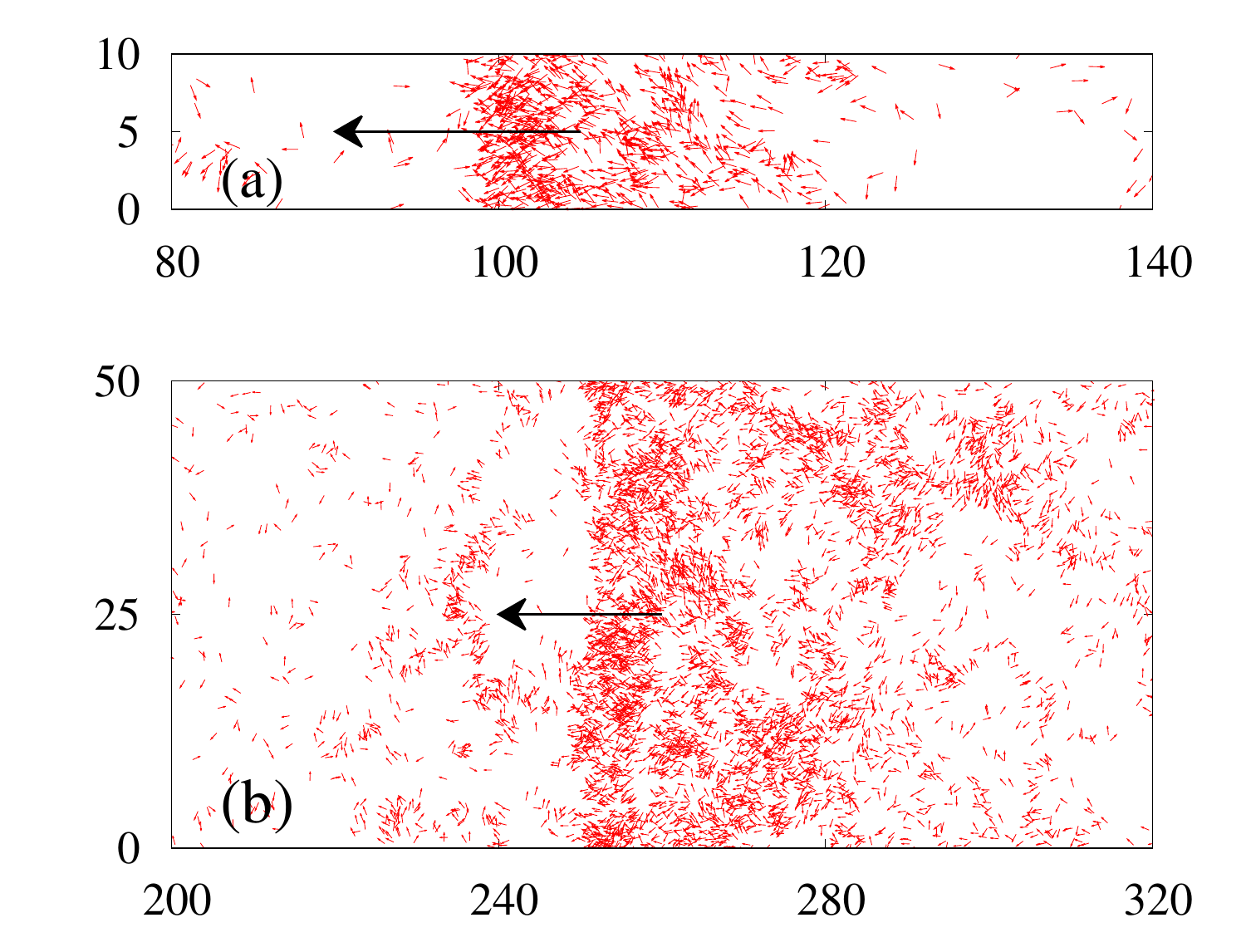}\\
\caption{Zoomed in snapshots (keeping $L_y$ fixed) showing propagation of a single band of a particular species for (a) $L_y=10$ (longitudinal density wave) and (b) $L_y=50$ (transverse density wave). The red arrows represent the orientation vector of the particles and the black arrows signify the direction of propulsion. Parameters: $L_x=400$, $\eta=0.3$, $\rho=1$ and $v_0=0.5$.} 
\label{fig12}
\end{figure}

Fig.~\ref{fig12} demonstrates the longitudinal and transverse motion directions of a high density liquid band in the coexistence regime. In Fig.~\ref{fig10}(a), we have seen that for $\eta=0.3$ and $\rho=1$, with $L_y=10$, band width $W>L_y$ whereas with $L_y=50$, $W<L_y$. When $W>L_y$, the band is both elongated (the width of the band moving in the $x$ direction is larger than its height $L_y$) and moving along the horizontal $x$ direction [Fig.~\ref{fig12}(a)] and this is a longitudinal density wave but for $W<L_y$, the band is elongated along the vertical $y$ direction but propels along the horizontal $x$ direction [Fig.~\ref{fig12}(b)] and this is a transverse density wave.

\end{document}